\documentclass[conference]{IEEEtran}

\usepackage[numbers]{natbib}
\setcitestyle{square}
\usepackage{graphicx}
\usepackage{color}
\usepackage{amssymb,amsmath,mathtools, mathrsfs, dsfont}
\usepackage{enumitem}
\usepackage{graphicx}
\usepackage{epstopdf}
\usepackage{mleftright}
\definecolor{blue}{rgb}{0,0,1}
\definecolor{darkgreen}{rgb}{0,.5,0}
\definecolor{darkred}{rgb}{.5,0,0}

\begin{document}
\title{Low Noise Non-Linear Equalization Using\\ Neural Networks and Belief Propagation}

\author{\IEEEauthorblockN{Etsushi Yamazaki, Nariman Farsad, Andrea Goldsmith\\
Department of Electrical Engineering Stanford University, Stanford, CA 94305\\
Email: etsushiy, nfarsad, andreag@stanford.edu}}
\date{October 2018}

\maketitle

\begin{abstract}
Nonlinearities can be introduced into communication systems by the physical components such as the power amplifier, or during signal propagation through a nonlinear channel. These nonlinearities can be compensated by a nonlinear equalizer at the receiver side. The nonlinear equalizer also operates on the additive noise, which can lead to noise enhancement. In this work we evaluate this trade-off between distortion reduction and noise-enhancement via nonlinear equalization techniques. We first, evaluate the trade-off between nonlinearity compensation and noise enhancement for the Volterra equalizer, and propose a method to determine the training SNR that optimizes this performance trade-off. We then propose a new approach for nonlinear equalization that alternates between neural networks (NNs) for nonlinearity compensation, and belief propagation (BP) for noise removal. This new approach achieves a 0.6 dB gain compared to the Volterra equalizer with the optimal training SNR, and a 1.7 dB gain compared to a system with no nonlinearity compensation.

\end{abstract}

\IEEEpeerreviewmaketitle

\section{Introduction}

\setlength\floatsep{3pt} 
\setlength\textfloatsep{3pt} 
\setlength\intextsep{3pt}
\setlength\abovecaptionskip{0pt}

In communication systems, many physical components have nonlinear responses, which distort signals traveling through them. For example, the power amplifier in radio transmitters, the Mach-Zender interferometer (MZI) modulator in optical communication systems, driving amplifiers for the MZI modulator, digital-analog/analog-digital converters, and the optical fiber channel itself are all sources of nonlinear distortion.

In these scenarios, digital signal processing can be used to equalize nonlinearity-induced distortions at the transmitter and at the receiver. When compensating for the nonlinearity at the transmitter, channel noise is not present, and hence, noise cannot be enhanced as part of equalization.  However, since the transmitter does not have any means to measure the distorted signal, additional feedback communication may be required for weight optimization. On the other hand, placing the nonlinearity equalizer at the receiver does not require feedback. For example, such an approach can compensate at the receiver for time-varying nonlinear distortion caused by a change in operation, temperature of the physical components, or stress on the fiber.

Nonlinear equalization at the receiver has to compensate for nonlinearities in the presence of noise, and this induces a noise enhancement problem. The noise enhancement in the digital filter is similar to that in nonlinear refractive index material, which is caused by four wave mixing in optical fiber \citep{90931} or modulation instability \citep{ZAKHAROV2009540}. More recently, digital back-propagation (DBP) \citep{Ip:08} has been well studied as a compensation method for optical fiber link nonlinearity. In particular, the noise enhancement in a DBP equalizer for fiber has been evaluated in \citep{Serena:16}. Thus, reducing the noise enhancement in the nonlinear equalizer is key to improving its performance. However, this can be a challenging problem since there is a trade-off between nonlinearity compensation and noise enhancement. 

To add to the complexity, the nonlinear response in the transmission system components can add a memory effect. Therefore, an algorithm that compensates for this memory effect is required. A common technique used for this compensation is the Volterra series approximation. A Volterra series approximation is a natural extension of the classical Taylor series approximation of linear systems to nonlinear systems. In the Volterra series approximation for the system output, in addition to the convolution of the input signal with the system's linear impulse response, the system output includes a series of nonlinear terms that contain products of increasing order of the input signal with itself. It can be shown that these polynomial extension terms allow for close approximations to the output for a large class of nonlinear systems, which basically encompasses all systems with scalar outputs that are time-invariant and have finite memory \citep{Schetzen}
. For this reason, the Volterra series approximation for the system output has been used in methods to compensate for the nonlinearity introduced by the transmitter components \citep{4538250zhu2008}, and by the fiber optical channel itself \citep{Shulkind:13}. Such methods are referred to as Volterra equalizers.

Recently, there have been many works on applying machine learning and neural networks (NNs) to digital communication systems.  For example, machine learning has been used for sequence detection \citep{8454325Nariman2018}, channel decoding of low-density parity-check (LDPC) codes \citep{8242643Nachmani2018}
, and joint source-channel coding \citep{8461983Nariman2018, DBLP:journals/corr/abs-1809-01733}.
Machine learning has also been used to compensate for the nonlinearity that may be introduced during communication. Some examples include compensating for the nonlinearity caused by transmitter clipping effects \citep{DBLP:journals/corr/abs-1809-01022} as well as those caused by the fiber channels itself \citep{Koike-Akino:18}. 
There are also several works that consider iterative integration of NNs with belief propagation (BP) channel decoding. In \citep{8259241Fei2018}, convolutional NN (CNNs) are used to remove correlated noise from the output of the BP decoder. In  \citep{DBLP:journals/corr/abs-1809-01022}, iterative decoding with NNs and LDPC codes is proposed, which helps to improve and compensate for severe inter-carrier distortion of orthogonal frequency division multiplexing (OFDM) signals caused by the transmitter clipping.

In this paper, we focus on nonlinearity compensation at the receiver. We first focus on a Volterra equalizer and derive an expression for the noise figure for distortion compensation that results in a common measure of signal-to-noise ratio (SNR) degradation. This derivation provides a mathematical expression for the noise enhancement effect of this equalizer, and supports our numerical analysis of the trade-off between nonlinearity compensation and noise enhancement. Using these results, we then propose a method to optimize the training SNR for the Volterra equalizer. We then propose an alternative approach to the Volterra equalizer that results in the best system performance. This approach alternates between NN equalization for compensation of the nonlinearity and BP for noise removal and channel decoding. In particular, we implement BP iterations as a nontrainable NN layer. This allows us to define the loss function that is used for training the NN equalizer as the cross entropy loss at the output of the BP step, which results in considerable gains in terms of performance. Finally, we evaluate the performance of this newly proposed approach, and demonstrate that it leads to a 1.7 dB gain versus no equalization, and outperforms the Volterra equalizer with the optimal training SNR by 0.6 dB. 

\section{System Description}

Figure 1(a) shows the digital transmission system assumed in this paper. The transmitter has a forward error correction (FEC) encoder, a bit-to-symbol mapping module, and a pulse shaping block. Let $M$ be the number of information bits per symbol for each of the in phase, $I$, and quadrature, $Q$, components of the signal. The binary information bits for the $m$-th bit of the $n$-th symbol at the FEC encoder output is denoted as $d_{n,m}$, and the amplitude for the $n$-th symbol at the bit-to-symbol mapper output is defined as $x_{n}$. Then, the pulse shaping block converts the single sample per symbol (SPS) waveform into multiple SPS. In this paper, 2 SPSs are used. 

\begin{figure}[!b]
\centering
\includegraphics[width=3.5in]{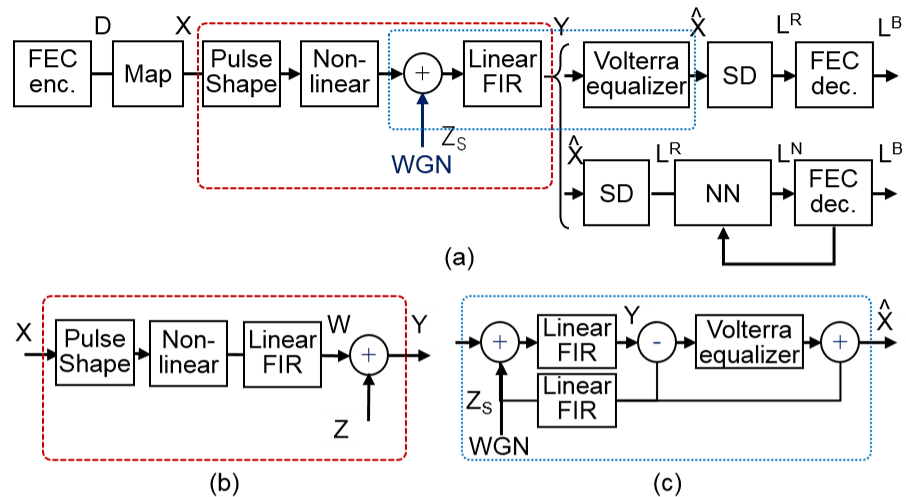}
\caption{Block diagram of the digital transmission system: (a) the model that is used for numerical simulations, (b) the approximate equivalent model to the red dashed blocks used for theoretical analysis of noise enhancement, (c) the replacement to the light blue dashed blocks where the noise is added after the nonlinear equalizer (i.e., noise is not enhanced by the Volterra equalizer). }
\label{fig:fig-system7png}
\end{figure}

The pulse shaped signal is then fed into a nonlinearity block, which captures the nonlinearity that could be introduced during transmission and signal propagation. This block could either introduce a memory-less nonlinearity or a nonlinearity with memory. In this paper, a memory-less nonlinearity is modeled as a sinusoidal transfer function given the simplicity of such functions as well as their generality in modeling nonlinearities in the system. Note that although the introduced nonlinearity is memoryless, when considered in combination with pulse shaping at the transmitter and the linear filter at the receiver, the nonlinear system response will have memory.

White Gaussian noise (WGN) is added to the signal in the channel, and at the receiver, a linear filter, a nonlinear equalizer, and a BP FEC decoder are used to recover the bits. The linear filter uses a pulse shaping function and an adaptive finite impulse response (FIR) filter to convert multiple-SPS signals into a single SPS signal. The amplitude level of the $n$-th symbol of the linear filter output is denoted as $y_{n}$. Another function of the adaptive FIR filter is to eliminate linear distortion. More precisely, this filter mitigates inter-symbol interference (ISI) by minimizing the mean square error (MSE) of its output.  In this paper, it is assumed that the pulse shaping functions of the transmitter and the receiver are both root raised cosine functions. Since the nonlinearity is modeled by the sinusoidal function, the total transfer function including the channel and linear filters is symmetric.

We consider two differed nonlinear equalization techniques: (A) A Volterra equalizer which is applied to the amplitude level signal, and (B) NNs with and without BP decoder feedback. The output amplitude levels of the Volterra equalizer and the soft decision (SD) inputs are denoted by $\hat{x}_{n}$. The outputs of the SD, NN, and BP FEC decoder have the form of a bit-wise log likelihood ratio (LLR), and they are denoted by $l^{\text{R}}_{n,m}$, $l^{\text{N}}_{n,m}$, and $l^{\text{B}}_{n,m}$, respectively. In SD, the output LLR is calculated as $l^{\text{R}}_{n,m}{=}\log{\{\Sigma_{x_{n}\in \chi_{m}^{0}}\exp({-}\rho|\hat{x}_{n}{-}{x}_{n}|^{2})\}}-\log{\{\Sigma_{x_{n}\in \chi_{m}^{1}}\exp({-}\rho|\hat{x}_{n}{-}{x}_{n}|^{2})\}}$, where $\rho$ is a parameter chosen based on the noise power, and $\chi_{m}^{0}$ and $\chi_{m}^{1}$ denote the sets of all possible ${x}_{n}$ for which $d_{n,m}{=}0$ or ${=}1$, respectively.

In the rest of the paper, the sequences of $x_{n}$, $y_{n}$, $\hat{x}_{n}$  are denoted by column vectors $\textbf{x} = \{x_{0},x_{1},...,x_{n}, ... \}^{\text{T}}$, $\textbf{y} = \{y_{0},y_{1},...,y_{n}, ... \}^{\text{T}}$, and $\hat{\textbf{x}} = \{\hat{x}_{0},\hat{x}_{1},...,\hat{x}_{n}, ... \}^{\text{T}} $, respectively.

Fig. 1(b) shows an approximated system setup to the blocks in the red dashed line in Fig. 1(a), which is used to derive analytic expressions for the noise figure in Section IV. In this setup, noise is added after the linear filter. The signal component of the output of the linear filter is denoted as W, and the noise components that is added to W is denoted as Z. The output of this equivalent approximation is denoted as $Y$ as in the original system setup.

Fig. 1(c) shows a system setup where the blocks inside the light blue dashed lines in Fig. 1(a) are replaced by those in Fig. 1(c). In this setup, the noise component, which is added in the channel, is removed just before the Volterra equalizer and is added again to the output of the Volterra equalizer. That is, the noise is added only after the Volterra equalizer. Therefore, there will be no noise enhancement by the Volterra equalizer. Note that in order to keep the operating conditions of the adaptive FIR filter exactly the same between the two systems, the FIR filter is applied to both the received noisy signal as well as to the noise component itself before the filter outputs are subtracted to remove the noise. We use this system setup in Section V to evaluate the noise enhancement in Volterra equalization.  

\section{Algorithms for Nonlinearity Equalization}

In this section, we discuss different techniques for nonlinear equalization. First, we describe the Volterra equalizer, which is the most widely-used currently. Then, we describe a new nonlinear equalization technique based on neural networks.

\subsection{Volterra Equalizer}

Since the Volterra equalizer considers the memory order of the nonlinearity, the output symbol of the Volterra equalizer is generated from consecutive input symbols $Y$ of the index range of $[n{-}L, n{+}L]$ with the $n$-th symbol at its center. Therefore, the filter length of the equalizer is defined as $2L{+}1$ symbols. Because here we assume a sinusoidal nonlinear transfer function as described in the previous section, which is odd symmetric, 2nd order product terms can be neglected. Thus, the input vectors required for the $n$-th output symbol generation are a set of 1st-order terms and 3rd-order product terms, which are row vectors of $ \textbf{y}^{(1)}_{n} $ and $ \textbf{y}^{(3)}_{n} $ given by
{\abovedisplayskip=5.0pt
{\belowdisplayskip=5.0pt
\begin{equation}
\begin{split}
\label{vltin}
\textbf{y}^{(1)}_{n} &= \big\{y^{(1)}_{n,i}\big\} {=} \big\{y_{n+i}| i{\in}[-L,L] \big\}, \nonumber \\
\textbf{y}^{(3)}_{n} &= \big\{y^{(3)}_{n,ijk}\big\} {=} \big\{y_{n+i}y_{n+j}y_{n+k}|(i,j,k) \in \mathbb{Z}_{L}^{3} \big\}, 
\end{split}
\end{equation}}}

\noindent where superscript $(1)$ and $(3)$ are used to refer to the terms being of 1st and 3rd order, respectively, and $\mathbb{Z}_{L}^{3}$ denotes the set of all possible combinations of 3 integers in the range of $[-L,L]$, defined as $\mathbb{Z}_{L}^{3} {=} \{(i,j,k)|i,j,k {\in} \mathbb{Z},~i{\in}[{-}L,L],~j{\in}[-L,i],~k{\in}[-L,j]\}$. The notation $ijk$ denotes the vector elements $(i,j,k)$. By defining the weights of the 1st and 3rd order product terms as column vectors $\textbf{h}^{(1)}$ and $\textbf{h}^{(3)}$, the output of the Volterra equalizer is given by the inner product of the weights and the input vectors as follows:
{\abovedisplayskip=0.0pt
{\belowdisplayskip=0.0pt
{\begin{equation}
\label{vltout}
\hat{x}_{n} {=} \textbf{y}^{(1)}_{n} \textbf{h}^{(1)}{+}\textbf{y}^{(3)}_{n} \textbf{h}^{(3)} {=} \sum_{i{=}-L}^{L}{h}^{(1)}_{i}{y}^{(1)}_{n,i}{+}\sum_{ijk{\in} \mathbb{Z}_{L}^{3}}{h}^{(3)}_{ijk}{y}^{(3)}_{n,ijk}.
\end{equation}}}}
All the input terms of the Volterra equalizer can be orthogonalized to each other, as was shown in \citep{319708Tseng}. Therefore, when $\textbf{h}^{(1)}$ and $\textbf{h}^{(3)}$ are optimized by minimizing the MSE between the estimated value and the transmitted value, $\hat{x}-x$, the optimum weights satisfy the following condition:
{\abovedisplayskip=5.0pt
{\belowdisplayskip=5.0pt
\begin{equation}
\begin{split}
\label{vltmse}
E_{n}[(\hat{x}_{n}-x) \cdot y^{(1)}_{n,i}] &= 0, \forall{i\in [-L,L]}, \\
E_{n}[(\hat{x}_{n}-x) \cdot y^{(3)}_{n,ijk}] &= 0, \forall{(i,j,k)\in \mathbb{Z}_{L}^{3}}. \\
\end{split}
\end{equation}}}
The sequence of the estimated symbols are given by $\hat{\textbf{x}} {=} \textbf{Y}^{(1)} \textbf{h}^{(1)} {+} \textbf{Y}^{(3)} \textbf{h}^{(3)} $, where two matrices defined by $\textbf{Y}^{(1)}{=}\{\textbf{y}^{(1)}_{0},\dots,\textbf{y}^{(1)}_{n},\dots\}^{\text{T}}$, and $\textbf{Y}^{(3)}{=}\{\textbf{y}^{(3)}_{0},\dots,\textbf{y}^{(3)}_{n},\dots\}^{\text{T}}$ are introduced. Using this representation, the optimum weights satisfying \eqref{vltmse} are given by
{\abovedisplayskip=5.0pt
{\belowdisplayskip=0.0pt
{\setlength\arraycolsep{2pt}
\begin{equation}
\label{vltwgt}
\left(
    \begin{array}{c}
        \textbf{h}^{(1)} \\
        \textbf{h}^{(3)}
    \end{array} 
\right) = \left[
    \begin{array}{cc}
        {\textbf{Y}^{(1)}}^{\text{T}}\textbf{Y}^{(1)} & {\textbf{Y}^{(1)}}^{\text{T}}\textbf{Y}^{(3)} \\ 
        {\textbf{Y}^{(3)}}^{\text{T}}\textbf{Y}^{(1)} & {\textbf{Y}^{(3)}}^{\text{T}}\textbf{Y}^{(3)}
    \end{array}
\right]^{-1} \left(
    \begin{array}{c}
        {\textbf{Y}^{(1)}}^{\text{T}}\textbf{x} \\ 
        {\textbf{Y}^{(3)}}^{\text{T}}\textbf{x} 
    \end{array}
\right).
\end{equation}}}}

\vspace{-.6cm}
\subsection{Neural network with belief propagation feedback}

In order to suppress the noise enhancement phenomena, a method to remove or cancel out white noise components of the input signal before or during the nonlinear equalization is required. One promising method to remove white noise is the BP decoding using the parity bit information. In SD decoding, the BP process is used iteratively to improve the LLR of each bit using those of other bits connected to identical check nodes. By successively using NNs with BP decoder steps, noise enhancement during nonlinear equalization is expected to be suppressed. This suppression of noise enhancement in nonlinear equalization leads to improvements in the input LLR at each BP step, and hence better performance. 
\begin{figure}[!b]
\centering
\includegraphics[width=3.5in]{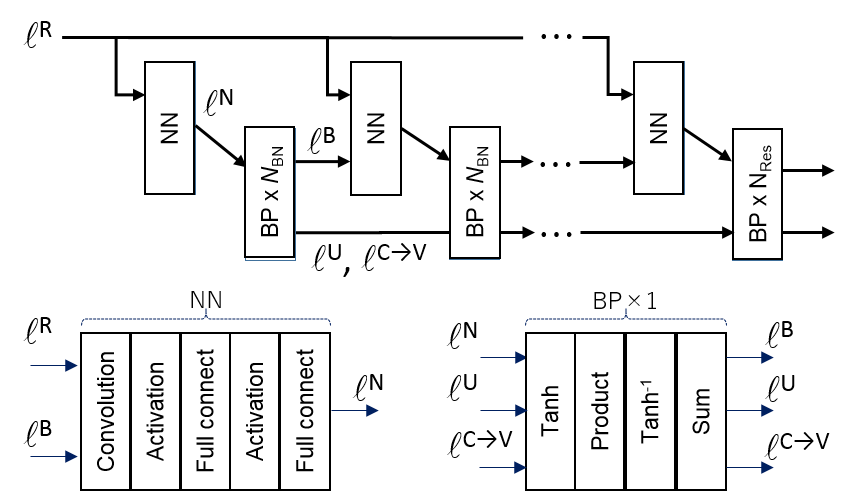}
\caption{Nonlinear equalization by iterative combination of NN and BP.}
\label{fig:fignnbp5.png}
\end{figure}

Fig.~2 shows the block diagram of nonlinear equalization operating with the BP decoder. 

The input to the NN equalizer has 2 streams of LLRs: the LLR after the linear filter $\textbf{l}^{\text{R}}$, and the LLR feedback from the previous BP decoder step $\textbf{l}^{\text{B}}$. The first NN stage is fed with only $\textbf{l}^{\text{R}}$. Then the LLR output from NN $\textbf{l}^{\text{N}}$ is fed into the BP decoder. The calculation flow of each stage of the NN is explained below. Since the goal is equalization of the nonlinearity with memory, the NN output LLR of the $n$-th symbol $l^{\text{N}}_{n,m}$ is generated by considering the input LLRs of the time window of $2L+1$ consecutive symbols centered on the $n$-th symbol:
{\abovedisplayskip=7.0pt
{\belowdisplayskip=7.0pt
\begin{equation}
\begin{split}
\label{nnin}
\textbf{\textit{l}}^{\text{R}}_{n} &= \{ l^{\text{R}}_{n+i,j}, i{\in}[-L,L], j{\in}[0,M-1] \} \nonumber\\
\textbf{\textit{l}}^{\text{B}}_{n} &= \{ l^{\text{B}}_{n+i,j}, i{\in}[-L,L], j{\in}[0,M-1] \} \nonumber\\
\end{split}.
\end{equation}}}
\noindent Here, integers $m$ and $n$ are used for the number of bit and symbol positions, respectively, and the same notation is used hereafter.
Since the input to the NN consists of consecutive symbols, a CNN is used in the first layer of the network as a sliding window that slides across the symbol sequence. The length of the window is designed depending on the memory length $L$. The output of the CNN layer is defined as ${q}_{n,m,k}$, where $k$ identifies the node in the next layer, and is given by:
{\abovedisplayskip=5pt
{\belowdisplayskip=0pt
{\setlength\arraycolsep{2pt}
\begin{align}
    q_{n,m,k} &= g_{1}(\sum\limits_{i{=}-L}^{L} \sum\limits_{j{=}0}^{M-1} w^{(\text{R})}_{m,k,i,j} \tanh(\frac{l^{\text{R}}_{n{+}i,j}}{2}), \nonumber \\ 
& \qquad \qquad {+} w^{(\text{B})}_{m,k,i,j} \tanh(\frac{l^{\text{B}}_{n{+}i,j}}{2}) {+} b^{(1)}_{m,k}),\nonumber
\end{align}
}}}
\noindent where $g_{1}$ is the activation function. 
Here, the function $tanh{(x/2)}$ is used to transform and limit the NN input variable range to within $[+1, -1]$. This is because the LLR value can range from negative to positive infinity, and additionally its probability distribution can vary depending on link conditions and the number of BP decoder iterations. 

The output of the first layer is then fed into a second  layer, which is a fully connected layer. 
The output of the second layer is defined as $r_{n,m,k} {=} g_{2}(\sum\limits_{j=0}^{N_{\text{q}}-1} w^{(2)}_{m,k,j} q_{n,m,j} {+} b^{(2)}_{m,k})$,
\noindent where $k$ denotes the node in the next layer, $N_{\text{q}}$ is the number of output nodes from the first layer, and $g_{2}(x)$ is the activation function.
Finally, the output layer is another fully connected layer which is given by
$l^{\text{N}}_{n,m} {=} g_{3}(\sum\limits_{j=0}^{N_{\text{r}}-1} w^{(3)}_{m,j} r_{n,m,j} {+} b^{(3)}_{m})$,
\noindent where $N_{\text{r}}$ is the number of outputs from the previous layer, and $g_3$ is the activation function.

In this paper, the ReLU function is used for $g_{1}(x) $ and $g_{2}(x)$, and linear activation is used for the final layer; $g_{3}(x){=}x$.

In order to train the NN and BP as a combined block, each iteration step of the the BP decoder is also implemented as a NN layer. At each step the BP receives three LLRs: an LLR from the NN $l^{\text{N}}_{n,m}$, the updated LLRs $l^{\text{U}}_{n,m}$, and $l^{\text{C} \rightarrow \text{V}}_{nm-k}$ from the previous iteration of the BP step. The output LLR of the BP step $l^{\text{B}}_{n,m}$ is calculated by summing $l^{\text{U}}_{n,m}$ and $l^{\text{N}}_{n,m}$, which is then passed to the next NN nonlinearity equalization step. Therefore, the BP step can be represented as
{\abovedisplayskip=5pt
{\belowdisplayskip=0pt
\begin{equation}
\begin{split}
\label{bpeq}
l^{\text{V}\rightarrow \text{C}}_{nm-k} &= l^{\text{N}}_{n,m} + l^{\text{U}}_{n,m} - l^{\text{C}\rightarrow \text{V}}_{nm-k}, \nonumber\\
l^{\text{C}\rightarrow \text{V}}_{nm-k} &= 2 \tanh^{-1}(\prod\limits_{(i,j)\in \mathbb{N}^{2}_{\text{V,C}(k)}\backslash{(n,m)}} \tanh(\frac{l^{\text{V} \rightarrow \text{C}}_{ij-k}}{2})), \nonumber\\
l^{\text{U}}_{n,m} &= \sum\limits_{k\in \mathbb{N}_{\text{C,V}(n,m)}} l^{\text{C} \rightarrow \text{V}}_{nm-k}. \nonumber\\
l^{\text{B}}_{n,m} &=  l^{\text{U}}_{n,m} + l^{\text{N}}_{n,m}. \nonumber\\
\end{split}
\end{equation}}}

We can use pairs of non-negative integers ($n$,$m$) to identify the variable node for the input LLR of the $m$-th bit of the $n$-th symbol, one of whose edge is connected to a check node of identification number $k$, as $\mathbb{N}^{2}_{\text{V,C}(k)}$. Similarly, we define a set of non-negative numbers for the check node, one of whose edge is connected to the variable node of identification pair number $(n,m)$, as $\mathbb{N}_{\text{C,V}(n,m)}$. Additionally, the LLR $l^{\text{C} \rightarrow \text{V}}_{nm-k}$ denotes a belief message passing from the check node of identification number $k$ to the variable node of $(n,m)$. Similarly, $l^{\text{V} \rightarrow \text{C}}_{nm-k}$ denotes a belief message passing from the variable node of $(n,m)$ to the check node of $k$. The BP decoder is repeated $N_{\text{BN}}$ times for each NN stage, and is repeated $N_{\text{Res}}$ times after the final NN stage.

The loss function is defined as a cross entropy between Tx bit-wise probability $p^{\text{T}}_{n,m}(d_{n,m})$ and bit-wise conditional probability of $d_{n,m}$ for a given BP output LLR $p^{\text{B}}_{n,m}(d_{n,m}|l^{\text{B}}_{n,m})$. Thus, the loss of the NN for the $m$-th bit $L_{m}$ is defined by ensemble over $n$ and summation over $d_{n,m}{=}{0,1}$ of ${-}p^{\text{T}}_{n,m}(d_{n,m})\log(p^{\text{B}}_{n,m}(d_{n,m}|l^{\text{B}}_{n,m}))$, and it is given by:
{\abovedisplayskip=5pt
{\belowdisplayskip=5pt
\begin{equation}
\label{loss}
L_{m} {=} E_{n}[ (1{-}d_{n,m})\log{(1{+}e^{-l^{\text{B}}_{n,m}})}{+}d_{n,m}\log{(1{+}e^{l^{\text{B}}_{n,m}})}],\nonumber
\end{equation}}
}
\noindent where $p^{\text{B}}_{n,m}(0|l^{\text{B}}_{n,m}){-}p^{\text{B}}_{n,m}(1|l^{\text{B}}_{n,m}){=}\tanh(l^{\text{B}}_{n,m}/2)$ is used.

The final loss is defined by the sum of all losses for all $m$ bits. In this paper, the modulation format of 64QAM is used, where each of the $I$ and $Q$ dimensions has 8 levels, and yields $M$=3. Since the most significant bit, $d_{n,m=0}$, is mapped to the sign of the corresponding symbol amplitude $x_n$, and its penalty of nonlinear distortion is negligible, the NN for the most significant bit is not implemented in this paper. Thus, NNs are configured for the second significant bit and the least significant bit only, and the final loss for each stage of the NN is given by $L = \lambda_{1} L_{1} + \lambda_{2} L_{2}$.

\vspace{-.05cm}
\section{Noise enhancement of nonlinear equalization}
\label{sec:noiseFig}
\vspace{-.1cm}
Noise enhancement can occur during nonlinear equalization. For example, let's assume that the input of the nonlinear equalizer has a signal component $w$ and a noise component $z$, with $w_{1}{+}z_{1}$, $w_{2}{+}z_{2}$, and $w_{3}{+}z_{3}$, where the noise power is much lower than the signal power. Then the 3rd order product $(w_{1}{+}z_{1})(w_{2}{+}z_{2})(w_{3}{+}z_{3})$ includes the pure signal term $w_{1}w_{2}w_{3}$, and all other terms have a noise component. Since we have assumed that the signal power is much larger than then noise power, the dominant noise terms will be $w_{1}w_{2}z_{3}$, $w_{1}z_{2}w_{3}$, and $z_{1}w_{2}w_{3}$.  Since each of these terms includes a product of noise and signal terms, they can be viewed as noise amplification or enhancement by the signal.

In this section, we analyze the approximate model shown in Fig.~1(b).  Recall that the input signal to the nonlinear filter $y_{n}$ consists of pure signal component $w_{n}$ and noise component $z_{n}$. Since we consider symmetric channels, and odd symmetric nonlinear response, the mean of the signal is zero, i.e. $E[w_{i}]{=}0$. We assume that the additive noise $z_{n}$ is zero mean, is independent for different symbols, i.e.,  $E[z_{i}z_{j}]{=}0$ for $i{\neq}j$, and is independent of signal, i.e., $E[w_{i}z_{j}]{=}0$ for any $i$, $j$. With these assumptions, the system model will be an approximation of the model in Fig. 1(a), since the linear filter can introduce dependence between $z_{i}$ and $z_{j}$ for $i{\neq}j$. This approximation allows us to analytically evaluate the noise enhancement in the Volterra equalizer and hence provides an analytical result to compare against our numerical simulation. By substituting $y_{n}$ in (\ref{vltout}) with $w_{n}{+}z_{n}$, the Volterra equalizer output is given~by
{\abovedisplayskip=0pt
{\belowdisplayskip=0pt
\begin{equation}
\begin{split}
\label{nfxhat}
\hat{x}_{n} &= \sum_{i=-L}^{L}{h}^{(1)}_{i}(w_{n+i}+z_{n+i})\nonumber\\
&+\sum_{ijk \in \mathbb{Z}_{L}^{3}}{h}^{(3)}_{ijk}(w_{n+i}{+}z_{n+i})(w_{n+j}{+}z_{n+j})(w_{n+k}{+}z_{n+k}).
\end{split}
\end{equation}
}}

The output can be separated into the compounded signal component $\hat{x}_{\text{S},n}$, which consists of only pure signal components of $w_{n}$, and the compounded noise component $\hat{x}_{\text{N},n}$, which includes only the terms that consist of the noise terms $z_{n}$. These compounded signal and noise terms are given by
{\abovedisplayskip=0pt
{\belowdisplayskip=0pt
{\abovedisplayskip=0.3pt
\begin{equation}
\begin{split}
\label{nfxsn}
  \hat{x}_{\text{S},n} & = \sum_{i=-L}^{L}{h}^{(1)}_{i}{w}_{n+i}{+}\sum_{ijk\in \mathbb{Z}_{L}^{3}}{h}^{(3)}_{ijk}{w}_{n+i} {w}_{n+j} {w}_{n+k}, \nonumber\\[-5pt]
  \hat{x}_{\text{N},n} & = \sum_{i=-L}^{L}{h}^{(1)}_{i}{z}_{n+i}{+}\sum_{ijk\in \mathbb{Z}_{L}^{3}}{h}^{(3)}_{ijk}\big( {w}_{n+i}{w}_{n+j}{z}_{n+k} \nonumber\\ 
  & \qquad + {w}_{n+i}{z}_{n+j}{w}_{n+k}+{z}_{n+i}{w}_{n+j}{w}_{n+k}+\dots \big).\nonumber
\end{split}
\end{equation}}}}
The signal power $P_{\text{S}}$ and the noise power $P_{\text{N}}$ of the Volterra equalizer output are given by the ensembles of their field squares, $P_{\text{S}}(\hat{x}){=} E[{\hat{x}}_{\text{S},n}^2]$  and $P_{\text{N}}(\hat{x}){=} E[{\hat{x}}_{\text{N},n}^2]$, respectively, as follows:
{\abovedisplayskip=0.pt
{\belowdisplayskip=0.pt
\begin{equation}
\begin{split}
\label{nfsn}
P_{\text{S}}(\hat{x}) &\cong \sum\limits_{i'=-L}^{L}\sum\limits_{i=-L}^{L}{h}^{(1)}_{i'}{h}^{(1)}_{i}E[{w}_{n+i'}{w}_{n+i}] \\
& {+}2\sum\limits_{i'=-L}^{L}\sum_{ijk\in \mathbb{Z}_{L}^{3}}{h}^{(1)}_{i'}{h}^{(3)}_{ijk}E[{w}_{n+i'}{w}_{n+i}{w}_{n+j}{w}_{n+k}], \\
P_{\text{N}}(\hat{x}) &\cong \sum\limits_{i'=-L}^{L}\sum\limits_{i=-L}^{L}{h}^{(1)}_{i'}{h}^{(1)}_{i}E[{z}_{n+i'}{z}_{n+i}] \\
&  {+}2\sum\limits_{i'=-L}^{L}\sum_{ijk\in \mathbb{Z}_{L}^{3}}{h}^{(1)}_{i'}{h}^{(3)}_{ijk}\big(E[{z}_{n+i'}{w}_{n+i}{w}_{n+j}{z}_{n+k}] \\
&  {+}E[{z}_{n+i'}{w}_{n+i}{z}_{n+j}{w}_{n+k}]{+}E[{z}_{n+i'}{z}_{n+i}{w}_{n+j}{w}_{n+k}]\big), 
\end{split}
\end{equation}
}}

\noindent where we have dropped the terms that included the square of $h^{(3)}_{ijk}$ in the assumption that $h^{(3)}_{ijk}E[w^{2}] {<<} h^{(1)}_{i}$, and this is reasonable considering that a ratio $E[w^{2}]|\textbf{h}^{(3)}| / |\textbf{h}^{(1)}|$
is observed to be 0.067 under the assumed nonlinearity.

We use the noise figure \cite{1695024Friis}
, which is defined as a ratio of the SNR at the input of the filter over the SNR at the output of the filter, to describe the noise enhancement feature of the nonlinear equalizer. Using \eqref{nfsn}, the noise figure $F$ for the Volterra equalizer is given by
{\abovedisplayskip=0.pt
{\belowdisplayskip=0.pt
{\setlength\arraycolsep{1pt}
\begin{equation}
\begin{split}
\label{nf1}
F \mspace{-3mu}=\mspace{-3mu}  \frac{P_{\text{S}}(y)/P_{\text{N}}(y)}{P_{\text{S}}(\hat{x})/P_{\text{N}}(\hat{x})} {=} \frac{|\textbf{h}^{(1)}|^{2}{+}2 \sum\limits_{i=-L}^{L}\sum\limits_{j=-L}^{i-1}\alpha_{i,j} \mathbf{\boldsymbol \Sigma}^{w11}_{i,j} }{{\textbf{h}^{(1)}}^{\text{T}} \mathbf{\boldsymbol \Sigma}^{w11} \textbf{h}^{(1)} {+} 2 {\textbf{h}^{(1)}}^{\text{T}} \mathbf{\boldsymbol \Sigma}^{w13} \textbf{h}^{(3)}}E[w^2] \\
\underset{j \leq i}{\alpha_{i,j}} = \sum\limits_{k=-L}^{j}{h}^{(1)}_{k}{h}^{(3)}_{ijk}{+}\sum\limits_{k=-L}^{i}{h}^{(1)}_{k}{h}^{(3)}_{ikj}{+}\sum\limits_{k=-L}^{L}{h}^{(1)}_{k}{h}^{(3)}_{kij} 
\end{split}
\end{equation}
}}}

\noindent where $P_{\text{S}}(y){=}E[w^2]$, $P_{\text{N}}(y){=}E[z^2]$, $\mathbf{\boldsymbol \Sigma}^{w11}{=} E_n[\textbf{w}^{(1)T}_{n}\textbf{w}^{(1)}_{n}]$, and $\mathbf{\boldsymbol \Sigma}^{w13}{=}E_n[\textbf{w}^{(1)T}_{n}\textbf{w}^{(3)}_{n}]$ are used. The notations $\textbf{w}^{(1)}_{n}$ and $\textbf{w}^{(3)}_{n}$ denote the signal component of $\textbf{y}^{(1)}_{n}$ and $\textbf{y}^{(3)}_{n}$, and they are defined by $\textbf{w}^{(1)}_{n} {=} \{w_{n+i}| i{\in}[-L,L] \}$ and $\textbf{w}^{(3)}_{n} {=} \{w_{n+i}w_{n+j}w_{n+k}|(i,j,k) {\in} \mathbb{Z}_{L}^{3} \}$ as row vectors, respectively. In the derivation of \eqref{nf1}, we used the fact that $E[w_{i}]{=}E[z_{i}]{=}E[w_{i}z_{j}]{=}0$ for $\forall i,j$ and $E[z_{i}z_{j}]{=}0$ for $\forall i,j,i{\neq}j$.

Note that the noise figure depends only on the statistics of the input signal, since the Volterra equalizer weights, $\textbf{h}^{(1)}$ and $\textbf{h}^{(3)}$ in \eqref{vltwgt}, depend on statistics of the input signal.

\section{simulation results}
In this section, we evaluate the performance of the nonlinear equalization techniques described in the previous sections. First, we use the analytical results on the noise enhancement properties of the Volterra equalizer to evaluate the trade-off between noise enhancement and nonlinearity reduction, and propose a new and better guideline for training the Volterra equalizer. We then compare the Volterra equalizer to our proposed approach: iterative NN equalization combined with BP noise removal. 

All our numerical simulations are performed using the system model shown in the Fig. 1(a). For the nonlinear equalizer, we use $L{=}4$ for both the Volterra and NN equalizers, and we use $N_{\text{q}}{=}N_{\text{r}}{=}40$ for the NN. We use a standardized DVB-S.2 LDPC channel codes with the code rate of 0.8 \citep{dvbs2}. The total number of iterations of the BP decoder is 50, with $N_{\text{BN}}=5$. For the linear filters, the adaptive FIR filter length of 17 is used, and the roll off factor of both the Tx and Rx pulse shapes is 0.2. 
\subsection{Nonlinear equalization and impact of white noise}

\begin{figure}[!b]
\centering
\includegraphics[width=3.4in]{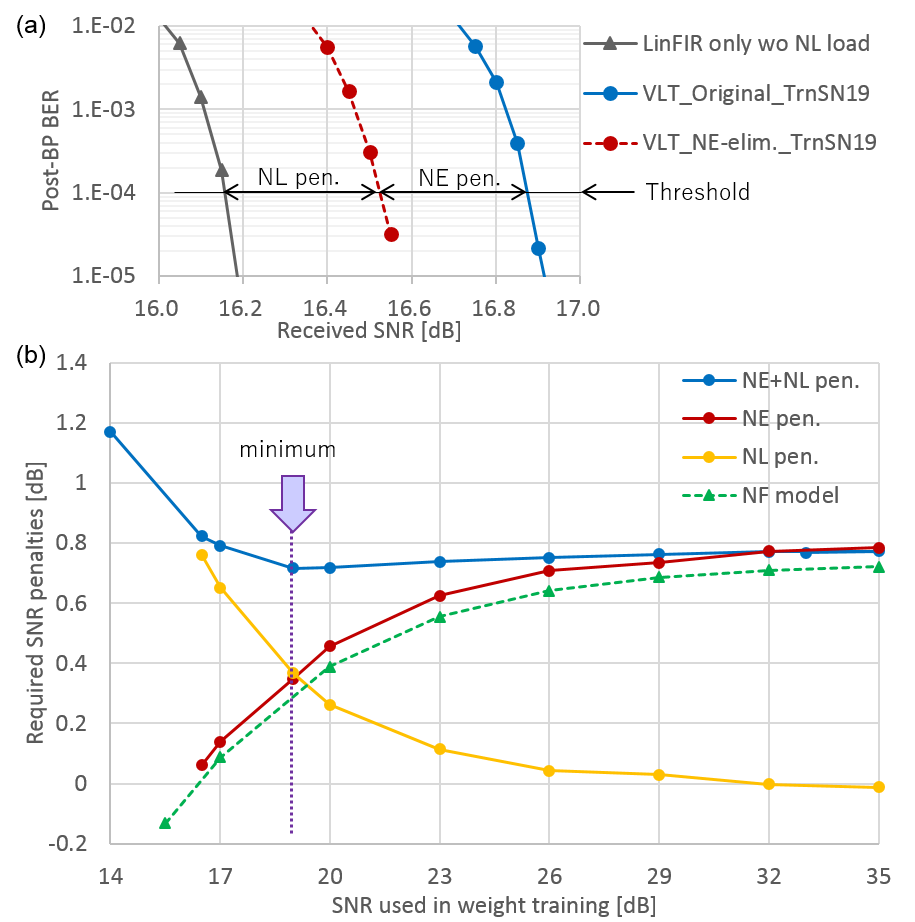}
\caption{(a) An example of post-BP BER vs SNR. The SNR used in training of Volterra (VLT) equalizer weights is 19 dB. (b) Noise enhancement penalty (NE-penalty), nonlinearity penalty (NL-penalty), and the total penalties as a function of the training SNR used for training the weights of the Volterra equalizer. The noise figure (NF) evaluated using \eqref{nf1} is plotted using the dashed green line. 
}
\label{fig:vlttrnsnr10.png}
\end{figure}

In order to demonstrate the noise enhancement property of the Volterra equalizer, we consider two systems. In particular, the systems in Fig. 1(a) and 1(c) are considered and compared. Recall that in Fig.~1(c) noise is added after the Volterra equalizer, and hence, the Volterra equalizer only compensates for the nonlinearity without enhancing the noise. 

Fig.~3(a) shows the result of this comparison. The black curve shows the BER of a system with no nonlinearity and no Volterra equalization. The blue curve on the right corresponds to the system in Fig.~1(a) where the nonlinearity is compensated by the Volterra equalizer. In this system the Volterra equalizer also enhances the noise. The red curve in the middle correspond to the system in Fig.~1(c), where the noise is added after the Volterra equalizer. Therefore, there is no noise enhancement by the Volterra equalizer; it only compensates for the nonlinearity. Thus, the difference between the black and the red curve corresponds to a nonlinearity penalty (NL-penalty), which cannot be removed by the Volterra equalizer, while the difference between the red and blue curve shows the Volterra equalizer noise enhancement penalty (NE-penalty).

In the rest of this subsection, we define the required SNR as the SNR required to achieve a post-BP BER performance of $10^{-4}$. Fig. 3(a) is one example of a post-BP BER plot, where the SNR that is used in training the Volterra equalizer weights is 19 dB. In this particular example, the NE-penalty, and the NL-penalty are observed to be 0.35 dB, and 0.37 dB, respectively, and their total is 0.72 dB.

\begin{figure}[!b]
\centering
\includegraphics[width=3.5in]{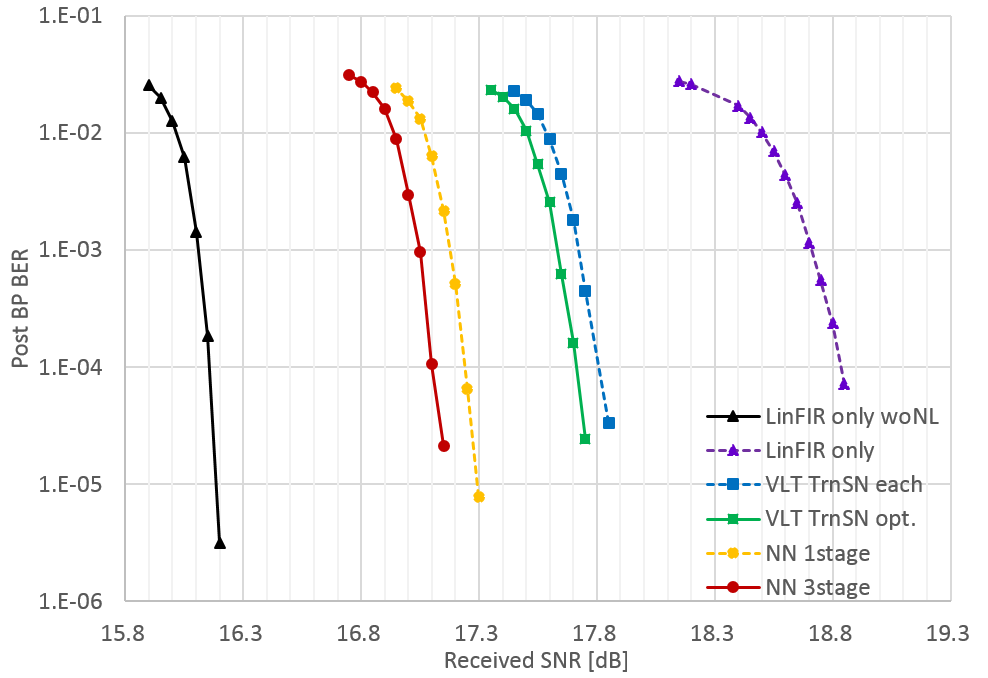}
\caption{Final comparison of FEC waterfall curves for Volterra and NN equalizer without/with BP decoder feedback. LinFIR: linear FIR, VLT: Volterra, woNL: without nonlinear load.}
\label{fig:fnlwaterfall3.png}
\end{figure}

Fig.~3(b) shows the NL-penalty, the NE-penalty, and their sum for various training SNR values. We thus see the trade-off between the NE-penalty and the NL-penalty as a function of the SNR that is used to train the weights of the Volterra equalizer. In particular, we observe that when a high SNR (e.g.,  35dB) is used to train the Volterra equalizer weights, it removes most of the signal distortion associated with the nonlinearity, at a cost of significantly enhancing the noise. Similarly, when a lower SNR (e.g.,  16.5 dB) is used for training the Volterra equalizer, the noise enhancement is reduced at the cost of not removing all the signal distortion associated with the nonlinearity. Interestingly, the total penalty remains relatively constant for training SNR values of more than 17dB, suggesting that it is best to train the Volterra equalizer at a high SNR. In Fig.~3(b), we also plot the noise figure derived in Section \ref{sec:noiseFig} (i.e., the dashed line). As can be seen, the derived noise figure is a good approximation to the NE-penalty of the Volterra equalizer. Finally, this figure demonstrates that there is a training SNR for the Volterra equalizer such that the total SNR penalty is minimized.

\subsection{Noise figure suppression by NN and BP}

In this subsection, comparison between the NN with BP feedback, which was proposed in section III, and the Volterra equalizer are presented based on simulations. 
Figure 4 shows the result for the post-BP BER as a function of the received signal SNR. The triangle solid line plot is a system with no nonlinearity. The triangle dashed line plot show the case where only the linear filter is used without any nonlinear equalization. Two results are shown for the Volterra equalizer: one where the Volterra equalizer is trained at each received SNR, and one where it is trained at the optimal SNR. We also consider two NN equalizers to compensate for the nonlinearity. First, we consider a single stage NN equalizer that is applied only once before BP feedback. Second a three stage iterative NN, where the first stage is before BP, and the second and third NN stages are after a few BP steps followed by more BP steps. As can be seen, the proposed iterative NN-BP equalization achieves the best performance with 0.6~dB gain compared to the best Volterra equalizer, and 1.7~dB gain compared to the case where there is no equalization to compensate for the nonlinearity.

\section{Conclusion}
We derived an analytic model of the noise figure for Volterra equalizers, which can be used to evaluate its noise enhancement. Using this model, the training SNR for the Volterra equalizer, which results in a better performance compared to training at each specific SNR, can be obtained.
Next, we proposed a new NN scheme for nonlinear equalization, where a BP step is implemented as a non-trainable NN layer, followed by another NN equalizer. This allows us to jointly decode and equalize to compensate for the nonlinearity. We show that the 3 stage NN equalizer with BP is better than the Volterra equalizer with optimal training SNR by 0.6~dB 
and has 1.7 dB gain compared to a nonlinear system with no nonlinearity compensation.

\bibliographystyle{IEEEtran} 
{\footnotesize

}

\end{document}